\documentclass[sigconf,nonacm]{acmart}

\setcopyright{none}
\acmDOI{}
    
\usepackage{xspace}
\usepackage{cleveref}
\usepackage{listings}
\usepackage{subcaption}
\usepackage{soul}
\usepackage{xcolor}

\usepackage{booktabs}
\usepackage{siunitx}
\usepackage[most]{tcolorbox}
\usepackage{xcolor} 
\usepackage{soul} 

\usepackage[normalem]{ulem}
\usepackage{textcomp}

\usepackage{balance}

\newcommand{\company}{{Google}\xspace}

\newcommand{\tool}{Cider Chat\xspace}

\newcommand{\mysec}[1]{\vspace{0.08cm} \noindent \textbf{#1}}

\usepackage{ifthen}
\newboolean{showcomments}
\setboolean{showcomments}{true} 
\ifthenelse{\boolean{showcomments}}
  {\newcommand{\nb}[2]{
    \fcolorbox{gray}{yellow}{\bfseries\sffamily\scriptsize#1}
    {\sf\small$\blacktriangleright$\textit{#2}$\blacktriangleleft$}
  }
  
  }
  {\newcommand{\nb}[2]{}
  
  }
  
\newcommand{\eg}{e.g.,\xspace}

\definecolor{added}{HTML}{AAFFAA}
\definecolor{deleted}{HTML}{FFAAAA}
\definecolor{edited}{HTML}{FFDDB3}

\lstdefinestyle{normal}{
  language=python,
  basicstyle=\ttfamily\scriptsize,
  aboveskip=0pt,
  belowskip=0pt,
  escapeinside={(*}{*)},
}

\lstdefinestyle{a}{
  language=python,
  basicstyle=\ttfamily\scriptsize,
  aboveskip=0pt,
  belowskip=0pt,
  backgroundcolor=\color{added},
  escapeinside={(*}{*)},
}

\newtcbtheorem{Summary}{\bfseries Summary}{enhanced,drop shadow={black!50!white},
  coltitle=black,
  top=0.15in,
  attach boxed title to top left=
  {xshift=1.5em,yshift=-\tcboxedtitleheight/2},
  boxed title style={size=small,colback=white}
}{summary}

\newtcolorbox[auto counter]{prompt}[1][]{title={\bfseries},enhanced,drop shadow={black!50!white},
  coltitle=black,
  top=0.1in,
  attach boxed title to top left=
  {xshift=1.5em,yshift=-\tcboxedtitleheight/2},
  boxed title style={size=small,colback=white},}

\begin{document}

\title{Reading Between the Lines: Scalable User Feedback via Implicit Sentiment in Developer Prompts}

\author{Daye Nam}
\authornote{Daye Nam conducted this research at Google.}
\affiliation{
  \institution{University of California Irvine}
  \country{}
}
\email{daye.nam@uci.edu}

\author{Malgorzata Salawa}
\affiliation{
  \institution{Google}
  \country{}
}
\email{magorzata@google.com}

\author{Satish Chandra}
\authornote{Satish Chandra is currently at Meta.}
\affiliation{
  \institution{Google}
  \country{}
}
\email{schandra@acm.org}

\begin{abstract}
Evaluating developer satisfaction with conversational AI assistants at scale is critical but challenging. User studies provide rich insights, but are unscalable, while large-scale quantitative signals from logs or in-product ratings are often too shallow or sparse to be reliable.
To address this gap, we propose and evaluate a new approach: using sentiment analysis of developer prompts to identify implicit signals of user satisfaction.
With an analysis of industrial usage logs of 372 professional developers, we show that this approach can identify a signal in \textasciitilde8\% of all interactions, a rate more than 13 times higher than explicit user feedback, with reasonable accuracy even with an off-the-shelf sentiment analysis approach.
This new practical approach to complement existing feedback channels would open up new directions for building a more comprehensive understanding of the developer experience at scale.
\end{abstract}

\maketitle

\section{Introduction}

The history of technology follows a recurring pattern: As breakthrough innovations mature from curiosities into essential infrastructure, the methods used to evaluate them must undergo an equally profound evolution. 
For example, early search engine evaluation focused on system-centric metrics like the size of its index, but the field's crucial insights came from analyzing user-centric signals derived from large-scale query logs, leading to many useful features that we still use in modern Web search~\cite{silverstein1999analysis, jansen2000real}.

Today, the rapid integration of Large Language Models (LLMs) into core developer workflows is at the same critical inflection point~\cite{jetbrainssurvey, sosurvey}.
As more users incorporate LLM tools into their workflows, more researchers and practitioners are investing effort in understanding the experiences of developers with these tools in real-world use~\cite{ziegler2022productivity, ziegler2024measuring, paradis2024how, chatterjee2024, cui2024}, and shifting the evaluation focus from static, model-centric benchmarks like HumanEval~\cite{chen2021evaluating} or MBPP~\cite{austin2021program}, toward a more holistic evaluation of understanding the benefits and harms of these tools within the real-world context~\cite{mozannar2024realhumaneval, jimenez2023swe}.

Initial efforts to understand the developer's perspective have relied on qualitative studies like interviews, surveys, and lab studies~\cite{paradis2024how, peng2023, kalliamvakou2022quantifying}. 
These works have provided invaluable early insights into a developer's perspective, such as perceived productivity, and have shaped our understanding of new innovations.
At the same time, these are still limited in that they are expensive and unscalable, often yielding highly focused insights with limited generalizability. 

To complement these approaches, in the field, especially researchers in industry, have turned to analyzing large-scale quantitative signals from real-world usage. 
This includes low-level engagement metrics like acceptance rate~\cite{ziegler2022productivity, dohmke2023sea} and, on the other end of the spectrum, high-level business metrics like user retention~\cite{jetbrainsretention}. 
However, although the acceptance rate has proven useful for simple interfaces like auto-completion~\cite{ziegler2024measuring}, it cannot support more recent, advanced tools like conversational AI, where users do not necessarily accept the suggestions. 
While user retention has the advantage of not being task-specific, it provides only a lagging, high-level signal; it tells us that a user was dissatisfied but provides no details as to why.
Thus, current large-scale approaches still do not provide an efficient way to analyze specific pain points or creative usages of real users to generate actionable product insights.

In theory, the ideal solution exists: direct, in-flow feedback, like thumbs-up/down ratings~\cite{copilotfeedback}. This mechanism is designed to capture user-driven feedback at scale, along with the in-situ context behind the interactions.
Yet, in practice, this promise goes unfulfilled.
The core issue is a fundamental lack of user motivation to interrupt their work, which results in an extremely sparse signal (0.6\% of interactions in our dataset; see \Cref{sec:results}). 

\begin{figure*}[hbt!]
    \centering
  \includegraphics[width=0.8\linewidth]{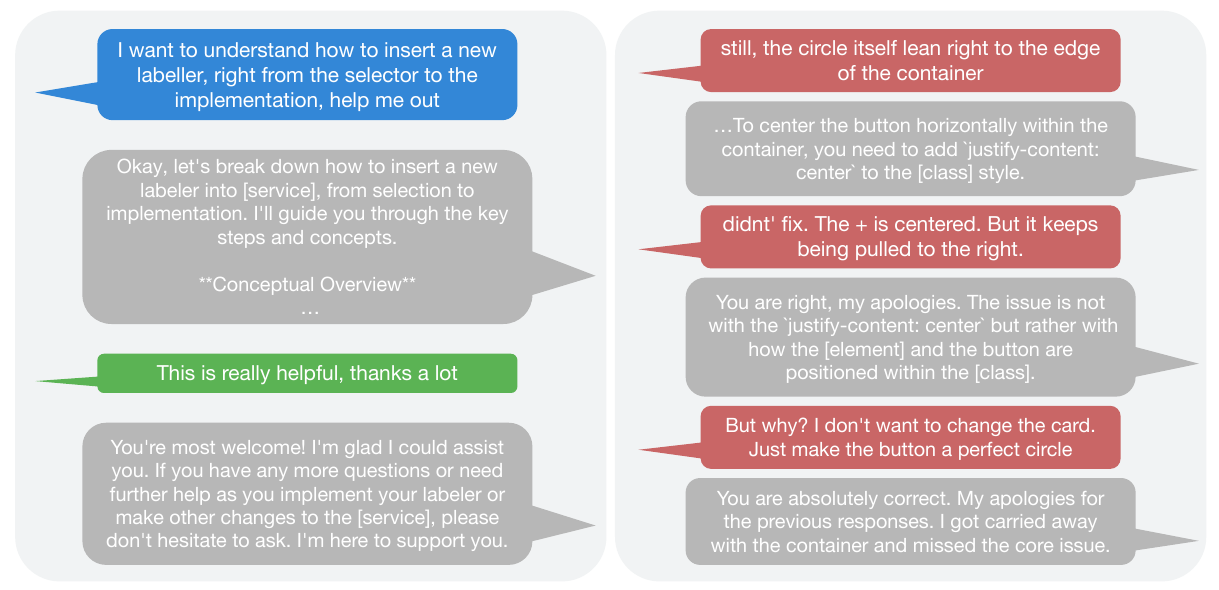}
  \caption{Example conversations containing positive (left) and negative (right) sentiment. User messages are shown in colored boxes (blue for neutral, green for positive, and red for negative sentiment), while AI responses are shown in grey boxes. }
  \label{fig:teaser}
  \vspace{-1em}
\end{figure*}

This led us to a different perspective: the key is not to ask users for feedback, but to listen to existing ones. This viewpoint led us to revisit the foundational paradigm of Computers Are Social Actors (CASA), first established in the 1990s~\cite{nass1994computers}. This theory posits that humans subconsciously apply social norms to computers. We hypothesized that this effect would be amplified with today's conversational developer tools, leading developers to naturally talk to the AI as a partner. This would make their language a rich, untapped source of implicit feedback, as illustrated in \Cref{fig:teaser}.

In this paper, we introduce and validate this new angle: \textbf{using automated sentiment analysis of developer prompts to serve as a scalable filter to discover insightful interactions}. 
We view this as a new capability for the scalable discovery of diverse and reliable user signals for further investigation, less as a comprehensive metric. Given the massive and rapidly growing volume of interactions, even a signal present in a fraction of conversations (\textasciitilde8\% in our study; see \Cref{sec:results}) can surface numerous actionable examples of user success and frustration, at a scale impossible to achieve through brute-force manual analysis.

To evaluate the feasibility and utility of this approach, we answer the following research questions (RQs):

\mysec{RQ1: Is the sentiment signal frequent enough to build a useful collection of examples?} We assess if our approach can surface a larger and more representative set of salient interactions compared to the existing thumbs-up/down feedback mechanism.

\mysec{RQ2: How accurately do automated sentiment scores reflect developer satisfaction?} We validate the precision of our filter against two ground-truth sources.

\mysec{RQ3: Do identified sentiments indicate longer-term product impact?} We test whether surfaced satisfactions or frustrations are merely anecdotal by examining whether a sentiment score correlates with user churn.

\section{Approach}
\label{sec:approach}

To investigate our research questions, we analyzed the telemetry logs of a coding agent in Cider, an internal IDE of \company that integrates advanced AI assistance, hereafter Cider Chat.
Similar to public tools such as GitHub Copilot Chat~\cite{copilot} or Cursor~\cite{cursor}, \tool assists developers by responding to natural language prompts that can be augmented with contextual information, such as highlighted code snippets. 
Its capabilities include answering questions about code, explaining errors, and brainstorming solutions.

\subsection{Data}
We analyzed usage data spanning from March 10 (Week 11) to May 11 (Week 19), 2025, from a pool of over 10,000 unique users. 
To focus on substantive interactions, we only included users who had made more than 10 requests to \tool in the given time frame, a threshold based on the median usage frequency. 
This is to minimize noise from trivial or exploratory interactions, such as new users testing the system with simple greetings.
From this population, we constructed our dataset by randomly sampling 372 users, which ensures a 95\% confidence level with a 5\% margin of error.
As the usage data are collected automatically on the server side, and as the data analysis was designed and performed independently, many biases~\cite{dell2012yours} that typically exist with lab studies were minimized.

\subsection{Sentiment Analysis}
We define a conversation as a complete thread of interaction between a developer and \tool. 
As shown in the examples in \Cref{fig:teaser}, each conversation is composed of multiple turns, with a user turn being the developer's prompt (colored boxes) and an AI turn being the agent's subsequent response (gray boxes).

Our core hypothesis is that a user's prompt often serves a dual purpose: it provides a new instruction while also implicitly conveying feedback on the quality of the preceding AI response. 
A positive follow-up prompt can signal a successful AI turn, whereas a frustrated or corrective prompt can indicate an unsatisfactory one. 
Therefore, we use the sentiment of a user turn as a potential quality metric for the AI turn that immediately precedes it.
Thus, we excluded the first user turn in any conversation from our sentiment analysis, as it has no preceding AI turn to react to. 
Similarly, conversations that consist of only a single user turn are omitted from our analysis.

For each qualifying user turn, our analysis involved a multi-step process. 
First, we used a general purpose LLM-based sentiment analysis prompt that has been designed and shared internally within \company for sentiment analysis tasks.
For a given text, this prompt returns a five-point sentiment category (from ``extremely negative'' to ``extremely positive'') along with its reasoning. In our experiments, we used Gemini 1.5 Flash. 
Text exceeding the model's context window was truncated, as long texts are usually due to the added context (\eg code) not influencing the user sentiment.

Second, because we observed that prompts consisting primarily of pasted error messages were frequently misclassified as negative (like in ~\cite{jongeling2017negative}), we implemented a refinement step. 
For any turn initially classified as negative, we used another LLM prompt to distinguish and extract the human-written portion of the raw error log. We then re-ran the sentiment analysis exclusively on this human-authored text.

Finally, we converted the categorical sentiment labels into numerical scores: -1.0 (extremely negative), -0.5 (negative), 0.0 (neutral), 0.5 (positive), and 1.0 (extremely positive). 
A conversation's overall sentiment score was calculated by averaging the scores of its qualifying turns. 

We acknowledge that there may be better sentiment analyzer and aggregation approaches, but we decided to adopt the simplest approach, as this is the first exploration of this idea. The prompts used can be found in the Appendix.

\subsection{Comparison with Direct User Feedback}
As a baseline for evaluating user satisfaction, we used the explicit feedback mechanism within \tool. 
The interface provides thumbs-up/down buttons for each AI response, allowing users to rate its utility directly. 
While this feedback is a direct signal of user satisfaction for a given turn, its usage is sparse.
In this analysis, we compare our sentiment-based approach against this explicit feedback to assess whether it provides a rich and accurate signal of user satisfaction.

\section{Results}
\label{sec:results}

\subsection{RQ1: Coverage and Applicability}

Our first research question examines the extent to which our sentiment analysis approach can be applied broadly, compared to existing feedback mechanisms. 
During the 9-week study period, the 372 sampled users engaged in \textasciitilde6,300 conversations with \tool, which comprised a total of \textasciitilde36,000 turns. 
This resulted in an average conversation length of 5.7 turns, with each user producing around 97 turns across 17 conversations on average.

\begin{figure}[t]
\centering
\includegraphics[width=\linewidth]{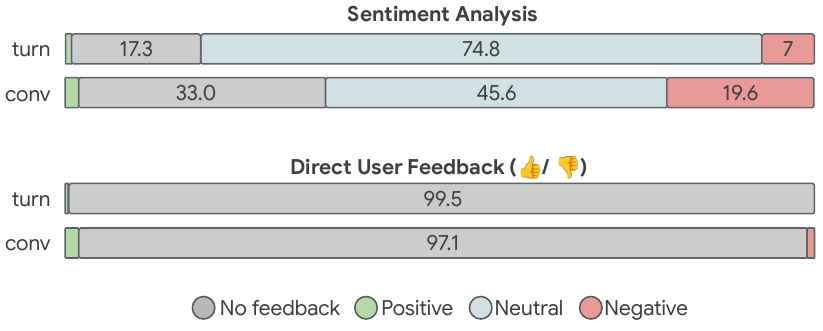}
\caption{Comparison of the proportion of turns and conversations via our sentiment-based approach versus direct user feedback. For sentiment analysis, no feedback indicates the first turn and the conversations that only contain a single turn. For direct user feedback, no feedback indicates the turns and conversations that did not receive any feedback. }
\label{fig:proportion}
\vspace{-1em}
\end{figure}

As described in \Cref{sec:approach}, sentiment analysis is applicable to any user turn that follows an AI response. As shown in \Cref{fig:proportion}, this means our approach can be applied to \textasciitilde83\% of all user turns (\textasciitilde29,900), with the remaining \textasciitilde6,300 turns being initial prompts. At the conversation level, this allows for evaluation of \textasciitilde67\% of all conversations (\textasciitilde4,200), excluding only those conversations that consisted of a single user turn.

Out of \textasciitilde83\% of the turns that were assigned sentiment scores,
\textasciitilde7\% were assigned negative sentiment scores, \textasciitilde1\% were assigned positive scores, and \textasciitilde75\% were neutral.
The extremely negative and extremely positive turns were the most interesting but sparse; we found
\textasciitilde{0.2}\% and \textasciitilde{0.03}\% respectively. 
Extremely positive turns included turns explicitly appreciating \tool, such as \textit{``Thanks, that fixed the build error and also wrapping the context in the location that you suggested worked great!''}.
Extremely negative turns, on the other hand, included deep frustration or, although rarely, even profanity words, 
like \textit{``this doesn't work, fix it now!!!!''}
The average sentiment score per user was -0.2, indicating that users are more prone to express negative sentiment than positive comments, which aligns with existing findings~\cite{jongeling2017negative}.
This could be because clarification or refinement, as in \Cref{fig:teaser}, is necessary to generate an improved response, while positive comments often are not.

In contrast, the direct user feedback mechanism (thumbs-up/down) yielded far sparser data. Feedback was provided for only \textasciitilde200 turns (0.6\%), corresponding to \textasciitilde120 unique conversations (2\%). This result confirms that our sentiment-based approach offers significantly wider coverage for identifying insightful user interactions.

\subsection{RQ2: Precision}

Our second research question assesses whether sentiment scores accurately approximate developer satisfaction. 
We answer this by a quantitative comparison with the sparse direct feedback data and a targeted manual verification.

\mysec{Quantitative Comparison with Direct User Feedback}
We first analyzed the turns in which both our sentiment score and a direct user feedback signal were present. After excluding turns with a neutral sentiment score, we identified 32 turns that were classified as either positive or negative by both methods. 

Despite the small sample size, a Chi-square test~\cite{mchugh2013chi} confirms a statistically significant association between the positivity of our sentiment score and the user's thumbs-up/down feedback ($\chi^2$(1,32)=7.98, p$<$0.01). This provides initial evidence that our metric aligns with direct user feedback.

\mysec{Manual Verification}
Given the sparsity of direct feedback, we performed a manual analysis to further validate our approach. We randomly sampled 50 user turns per category.
Two authors of this paper then independently labeled each turn as indicating positive, negative, or neutral sentiment, in the context of the conversation, including the previous and next user turns and AI responses.
The protocol used for manual labeling can be found in Appendix.

The initial agreement between the two human raters was high, with a Cohen's Kappa of 0.72, indicating substantial agreement.
Disagreements were resolved through discussion. 
We then compared our automated sentiment scores to this manually-created ground truth. 
We found that our method's sentiment label matched the final human label in 75\% of the cases, with 69\%, 75\%, and 85\% matches on the positive, neutral, and negative classes, respectively.
This confirms that even simple sentiment analysis can be a reasonable and promising proxy for developer satisfaction, which could save a lot of time in manual review of usage data.

\subsection{RQ3: Association with Longer-term Impact}

For our final research question, we explore whether aggregated sentiment can serve as a signal related to long-term user engagement. We acknowledge that a developer's decision to adopt, continue, or abandon a tool is complex and is influenced by many factors (e.g., tool performance)~\cite{larios2020selecting, witschey2015quantifying}. Therefore, our goal is not to build a predictive model of churn, but to investigate if the prompt sentiment shows a plausible association with user retention as an early signal. We hypothesize that a sustained negative experience will be associated with a lower likelihood of continued use.

We used two periods from our dataset: an initial period (Weeks 12-15) and a subsequent period (Weeks 16-19). 
For each user, we calculated their average sentiment score from the initial period. 
We defined a binary variable, {\small \texttt{did\_return}}, which was true if the user had at least one request to \tool in the subsequent period, and false otherwise.

\begin{figure}[t]
    \centering
    \includegraphics[width=0.6\linewidth]{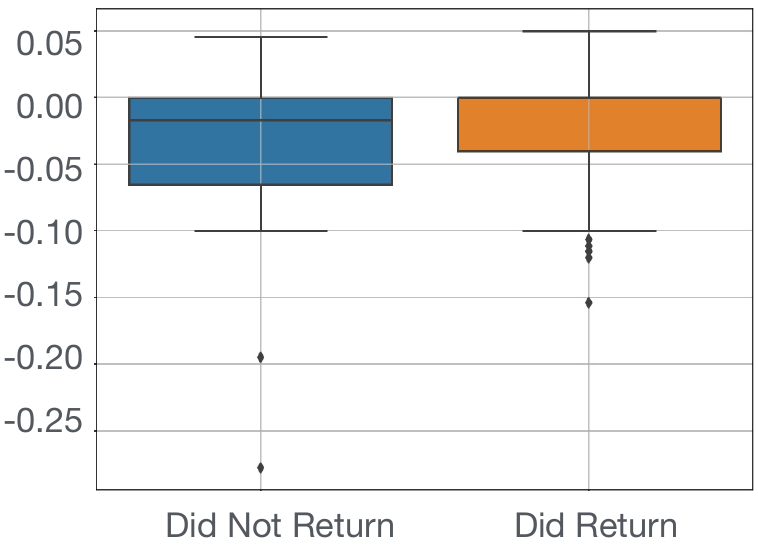}
    \caption{Boxplots of average sentiment scores (week 12-15) for returning vs. non-returning users (week 16-19).}
    \label{fig:abandonment}
    \vspace{-1em}
\end{figure}

A Point-Biserial correlation test~\cite{tate1954correlation} revealed a statistically significant, albeit modest, positive correlation between a user's average sentiment score in the initial period and their likelihood of returning in the subsequent period (coeff = 0.15, p$<$0.01). 
We note that this does not provide very strong results, as this correlation does not imply causation, and the coefficient of 0.15 only indicates low correlation.
However, the direction of the relationship is consistent with our hypothesis.
As visualized in \Cref{fig:abandonment}, users who returned to the tool had a distribution of sentiment scores concentrated around neutral values (median $\sim$ 0), while users who did not return exhibited a wider, more negative distribution.
While there are many other factors that can impact the user churn, which would require more robust future study, this may indicate that aggregated sentiment may serve as a valuable, automated signal for monitoring user experience at scale and identifying segments of users who might be having a negative experience.

\section{Discussion}
In this section, we discuss the practical implications of our findings and acknowledge the limitations of our approach.

\subsection{Implications and Use Cases}

\mysec{Operationalizing Sentiment for Insight Discovery} 
The most immediate application of our approach is to operationalize the sentiment signal as a scalable discovery mechanism for in-depth user research. Automatically flagging turns with strong negative or positive sentiment provides a rich, contextualized stream of data for error analysis, helping teams identify unexpected failure modes without the need for manual searching. Furthermore, researchers can move beyond individual interactions by using aggregate sentiment scores to identify and recruit specific user segments, such as the most satisfied or most frustrated users, for deeper qualitative studies to understand their core pain points and best practices.

\mysec{Curating High-Impact Training Data}
Beyond user research, our approach can also be useful in addressing a critical bottleneck in model development: the curation of high-quality training data. Manually finding salient examples of success or failure for supervised fine-tuning is infeasible given the sheer volume of interaction logs and the need for manual review. By automatically filtering for high-sentiment and low-sentiment conversations, engineering teams can efficiently build focused datasets that reflect real-world user challenges and successes, enabling more targeted and effective model improvements.

\mysec{Understanding User Struggle at Scale} 
Finally, our validated sentiment score can be used as a proxy for ground-truth user satisfaction to discover and validate other behavioral signals of user struggle. By correlating sentiment with interaction metadata, we can identify which user actions are indicative of frustration.
For example, we found that conversations with negative sentiment tend to be significantly longer (\textasciitilde10 turns) than positive (\textasciitilde6 turns) or neutral (\textasciitilde4 turns) ones (see \Cref{fig:length} in Appendix). This indicates conversation length as one such quantitative signal for this common pain point. This same approach could be used to investigate other potential signals, such as code churn or high response regeneration rates, allowing them to be utilized even when users do not express explicit sentiment.

\subsection{Limitations and Threats to Validity}
While promising, we recognize several limitations to our work that should be taken into account.

\mysec{Perception vs. Ground Truth}
The sentiment scores are based on the user's perception of AI responses, not the objective correctness. 
A user might be frustrated by a correct response that they do not understand, or satisfied with a plausible but subtly incorrect one. 
This is a fundamental challenge for any feedback-based evaluation.

\mysec{Confounding Variables} 
A user's expressed sentiment is influenced by many factors, such as their personality or stress level.
Therefore, we do not claim that it can be an absolute, standalone metric, but it should be used as one of many signals for usage analysis.

\mysec{Limited Generalizability \& Transparency}
As our analysis relies on internal usage data from \company, we cannot share the raw data. However, given the scale of the dataset, we believe that the high-level findings may generalize to other contexts.

\section{Future Plans}
Our future work will proceed along two parallel tracks:

\mysec{Strengthening the Sentiment Analysis}
This paper demonstrates the feasibility of our approach using a general-purpose, off-the-shelf sentiment analyzer. Our immediate future work will focus on strengthening this core signal. We plan to develop a domain-specific sentiment model, optimized for developer terminology and cost-effective for large-scale deployment. Furthermore, we will explore more sophisticated aggregation methods beyond simple averaging, such as weighting by sentiment intensity, to better model how sentiment reflects longer-term user perception and trust.

\mysec{Understanding the AI-Powered Developer Experience}
A robust sentiment signal will enable a wider range of studies to build a more holistic understanding of the AI-powered developer experience. We plan to conduct large-scale correlational studies linking sentiment with other objective developer behaviors (\eg change throughput, active coding time) and user characteristics (\eg programming experience, AI familiarity). This will enable us to triangulate and verify findings from smaller-scale qualitative studies, testing open research questions about developer productivity and satisfaction on a larger scale.

\newpage

\balance
\bibliographystyle{ACM-Reference-Format}
\bibliography{references}

\onecolumn
\appendix

\section{System instruction for sentiment analysis}
\begin{prompt}
You are an expert at analyzing the sentiment of text.
Given a piece of text, rate the sentiment
of the text on a scale ranging from extremely negative, negative, neutral,
positive, to extremely positive.

We want to analyze text based on the given label definitions.
\\

Labels definitions:
\begin{itemize}
    \item extremely negative: Texts that use extremely negative language or emojis to express intense negative emotions such as hate, disapproval, dissatisfaction , frustration. The overall tone is strongly critical and can be very sarcastic.
    \item negative: Texts that use moderately negative language to express dissatisfaction, disapproval or discomfort without using extreme language.  The overall tone is critical to some extent but tends to be more restrained and less emotionally charged compared to stronger negative sentiments.
    \item neutral: Texts that lack positive or negative emotions and are typically objective, factual, or balanced, without conveying a clear bias or emotional tone. Text is often straightforward statement of preference or curiosity without expressing strong approval, disapproval, enthusiasm, or dissatisfaction. Additionally, neutrality doesn't necessarily imply a complete absence of emotion but rather a lack of strong, identifiable positive or negative emotional inclinations.
    \item positive: Texts that use moderately positive language to express satisfaction, approval or admiration but are not overly enthusiastic or intense. Usually adjectives similar to pleasant, satisfactory, encouraging, optimistic, amiable, express mildly positive sentiment.
    \item extremely positive: Texts that use strongly positive language to express emotions such as extreme enthusiasm, satisfaction, or admiration or use happy emojis. Identify the positive adjectives used and their degree to predict the sentiment. Usually positive adjectives of superlative/comparative degree express super positive sentiment.\\
\end{itemize}

Here is the text to analyze:

[[text]]\\

Think about a rationale step-by-step for making a sentiment scale choice
before doing so. Include an explanation of your reasoning. Restrict your response to two lines.
\end{prompt}

\section{System instruction for human message separation}
\begin{prompt}
You are an expert software engineer specializing in parsing
and categorizing technical text. Your task is to analyze the provided input text
 and separate any human-written messages from computer-generated error messages
(e.g., compiler errors, stack traces, runtime exceptions).\\

    Labels definitions:
    \begin{itemize}
        \item error\_message\_only: The text consists solely of an error message.
        \item human\_message\_only: The text consists solely of a human-written message.
        \item mixed\_message: The text contains both a human-written message and an error message.\\
    \end{itemize}

    Separation guide:
    \begin{itemize}
        \item If the classification is mixed\_message, identify and extract the human-written message.
        \item If the classification is error\_message\_only, indicate an empty string for the human message.
        \item If the classification is human\_message\_only, extract the human message.\\
    \end{itemize}

Think about a rationale step-by-step.
Include an explanation of your reasoning. Restrict your response to two lines.\\

Here is the text to analyze:

[[text]]
\end{prompt}

\newpage

\section{Manual Annotation Protocol for User Satisfaction}

The primary objective of manual annotation is to capture the \textit{user's perceived satisfaction}, not the objective correctness of the AI's output, which is extremely challenging to evaluate without having a full understanding of user intent and the task. 
A label will be assigned to a set of sampled user turns ($n^{th}$ turn in conversation) which indicated satisfaction with the previous AI response ($(n-1)^{th}$ turn in conversation). 

\subsection{Annotation Process}
For each sampled user turn ($n^{th}$ turn), the annotation can follow these steps:
\begin{enumerate}
    \item \textbf{Examine the Target Turn:} First, analyze the content of the user's prompt at the $n^{th}$ turn. If the user's satisfaction or dissatisfaction is explicitly stated or strongly implied, assign a label.
    \item \textbf{Examine Surrounding Context:} If the target turn is ambiguous, expand the context to include:
    (1) the user's previous prompt ($(n-2)^{th}$ turn),
    (2) the AI's response that the user is reacting to ($(n-1)^{th}$ turn),
    (3) the AI's next response ($(n+1)^{th}$ turn), as it sometimes contains clues (e.g., an apology).
    \item \textbf{Assign a Label:} Based on the evidence, assign one of the three labels defined below. To maintain efficiency, if a definitive label could not be assigned within approximately 3 minutes of analysis, the turn was labeled `Cannot Judge`.
    \item \textbf{Handle Non-English Text:} Any prompts written in languages other than English were translated before labeling.
\end{enumerate}

\subsection{Label Definitions and Heuristics}

\mysec{Satisfied}
A ``Satisfied'' label can be assigned if the user's prompt indicated that the previous AI response met their needs or made successful progress on their task. Turns with following characteristics can be assigned ``Satisfied'':
\begin{itemize}
    \item Explicit appreciation (e.g., \textit{``Great!''}, \textit{``Awesome''}, \textit{``thanks!''}).
    \item Smooth continuation to a subsequent step (e.g., ``Perfect, now can you add comments to that function?'').
    \item Affirmation followed by a new, related request.
\end{itemize}

\mysec{Unsatisfied}
An ``Unsatisfied'' label was assigned if the user's prompt indicated that the previous AI response was incorrect, unhelpful, or failed to meet their expectations.
Turns with following characteristics can be assigned ``Unatisfied'':
\begin{itemize}
    \item Explicit frustration or negative feedback (e.g., \textit{``No, that's wrong,''} \textit{``This is useless.''}).
    \item Corrective statements pointing out an error in the AI's previous response (e.g., \textit{``No! You forgot to handle the null case.''}).
    \item A follow-up question phrased as a complaint or pointing out a flaw (e.g., \textit{``Why did you use a deprecated library?''}).
    \item The AI's subsequent response is an apology (e.g., if the AI says \textit{``I apologize for the confusion,''} it is highly likely the preceding user turn expressed dissatisfaction).
    \item The user encounters a new error that was directly caused by the AI's last suggestion.
\end{itemize}

\mysec{Cannot Judge}
This label was used when there was insufficient evidence in the conversation to confidently infer the user's satisfaction level.
Turns with following characteristics can be assigned ``Cannot Judge'':
\begin{itemize}
    \item The user's prompt is a simple statement of a new task or error with no emotional or qualitative language (e.g., a user pastes in a new error message without comment).
    \item The user's prompt is an unrelated question, indicating a complete context shift from the previous turn.
    \item The intent is simply too ambiguous to label reliably within the allotted time.
\end{itemize}

\newpage

\section{Additional figures}

\begin{figure}[htb]
\centering
\includegraphics[width=0.6\linewidth]{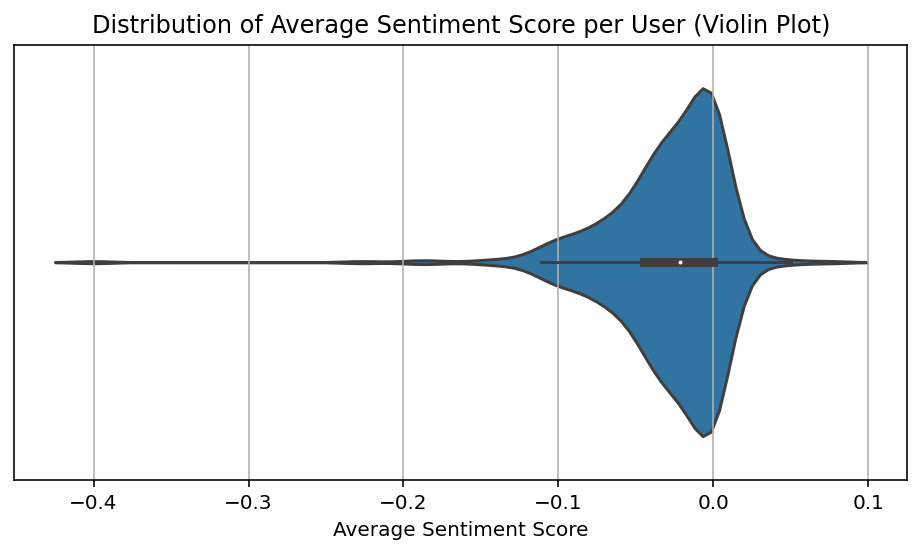}
\caption{Average sentiment score per user.}
\label{fig:peruser}
\end{figure}

\begin{figure}[htb]
    \centering
    \includegraphics[width=0.6\linewidth]{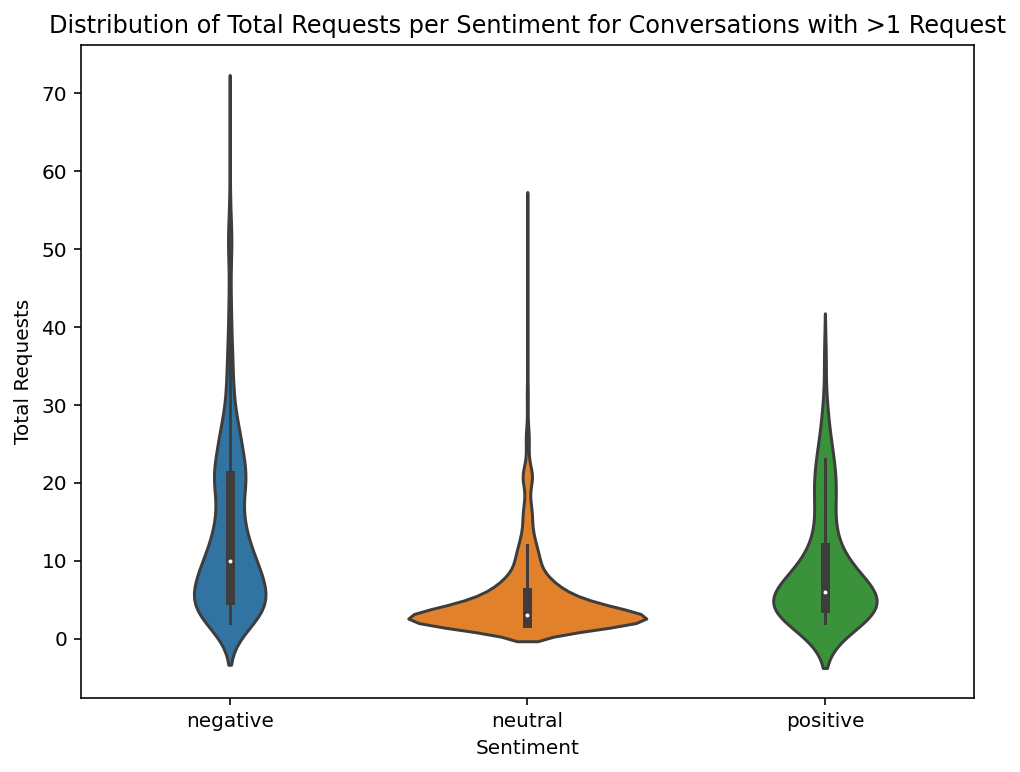}
    \caption{Distribution of the number of turns included in each conversation, grouped by the conversation sentiment.}
    \label{fig:length}
\end{figure}

\end{document}